\documentclass[a4paper,10pt,twoside]{cpc-hepnp}
\usepackage{upgreek,fancyhdr}
\usepackage{multicol}
\usepackage{graphicx}
\usepackage{booktabs}
\usepackage{amssymb,bm,mathrsfs,bbm,amscd}
\usepackage[tbtags]{amsmath}
\usepackage{lastpage}
\graphicspath{{figures/}}
\usepackage{subfigure}
\usepackage{url}
\usepackage{dcolumn}
\usepackage{CJKutf8}

\begin{document}
\begin{CJK*}{UTF8}{gbsn}

\fancyhead[c]{\small Chinese Physics C  Vol. 40, No. 5 (2016) 057004}
\fancyfoot[C]{\small 057004-\thepage}
\footnotetext[0]{Received 10 October 2015}

\title{A new muon-pion collection and transport system design using superconducting solenoids based on CSNS
\thanks{Supported by National Natural Science Foundation of China (11527811)}}

\author{ Ran Xiao (肖冉)$^{1,2}$\quad Yan-Fen Liu (刘艳芬)$^{1,2}$ 
\quad Wen-Zhen Xu (许文贞)$^{3}$\quad Xiao-Jie Ni (倪晓杰)$^{1,2}$ \\
\quad Zi-Wen Pan (潘子文)$^{1,2}$\quad Bang-Jiao Ye (叶邦角)$^{1,2;1)}$\email{bjye@ustc.edu.cn}
}

\maketitle

\address{%
$^1$ State Key Laboratory of Particle Detection and Electronics, University of Science and Technology of China, Hefei 230026, P. R. China\\
$^2$  Department of Modern Physics, University of Science and Technology of China, Hefei 230026, P. R. China\\
$^3$ Shanghai Synchrotron Radiation Facility, Shanghai Institute of Applied Physics, Chinese Academy of Sciences, Shanghai 201204, China
}

\begin{abstract}
A new muon and pion capture system is proposed for the China Spallation Neutron Source (CSNS), currently under construction. Using about 4\% of the pulsed proton beam (1.6 GeV, 4 kW and 1 Hz) of CSNS to bombard a cylindrical graphite target inside a superconducting solenoid, both surface muons and pions can be acquired. The acceptance of this novel capture system - a graphite target wrapped up by a superconducting solenoid - is larger than the normal muon beam lines using quadrupoles at one side of the separated muon target. The muon and pion production at different capture magnetic fields was calculated using Geant4. The bending angle of the capture solenoid with respect to the proton beam was also optimized in simulation to achieve more muons and pions. Based on the layout of the muon experimental area reserved at the CSNS project, a preliminary muon beam line was designed with multi-purpose muon spin rotation areas (surface, decay and low-energy muons). Finally, high-flux surface muons (10$^8$/s) and decay muons (10$^9$/s) simulated by G4beamline will be available at the end of the decay solenoid based on the first phase of CSNS. This collection and transport system will be a very effective beam line at a proton current of 2.5 $\mu$A. 
\end{abstract}

\begin{keyword}
surface muon; decay muon; large acceptance channel; superconducting solenoids
\end{keyword}

\begin{pacs}
29.27.Eg, 29.27.Fh, 85.25.Am
\end{pacs}

\footnotetext[0]{\hspace*{-3mm}\raisebox{0.3ex}{$\scriptstyle\copyright$}2013
Chinese Physical Society and the Institute of High Energy Physics
of the Chinese Academy of Sciences and the Institute
of Modern Physics of the Chinese Academy of Sciences and IOP Publishing Ltd}%

\begin{multicols}{2}

\section{\label{sec1}Introduction}
Muon spin rotation/relaxation/resonance ($\mu$SR) is a useful local probe technique to investigate the chemical and physical properties of condensed matter. It provides complementary information to other similar methods, such as nuclear magnetic resonance (NMR) and Mossbauer spectroscopy, because of its unique characteristics: extreme sensitivity to small internal magnetic fields, the detection of magnetic fluctuations over a large range (10$^4\sim 10^{12}$/s) and applications in a large variety of sample environments \cite{Yaouanc2011}. Surface muons are generated from low-energy pions which have lost their whole momentum inside the production target and stop at the target surface layer. These positive pions decay at rest and produce monochromatic ($\sim$29.8 MeV/c) muons with nearly 100\% spin polarization \cite{nagamine2003}. The stopping range of the surface muon is about 150 mg/cm$^2$ and the range straggling is within a layer of 20 mg/cm$^2$ \cite{brewer1981}. This high luminosity is a great advantage and allows $\mu$SR measurements of materials. Decay muons are obtained from the decay of pions in flight. They are another kind of muon source with higher momentum ($>$ 60 MeV/c) and lower spin polarization (about 80$\sim$90\%). The collection and transport system of pions is very important for a high intensity of decay muons.

\begin{figure*}[t]
\centering
\includegraphics[width=10 cm]{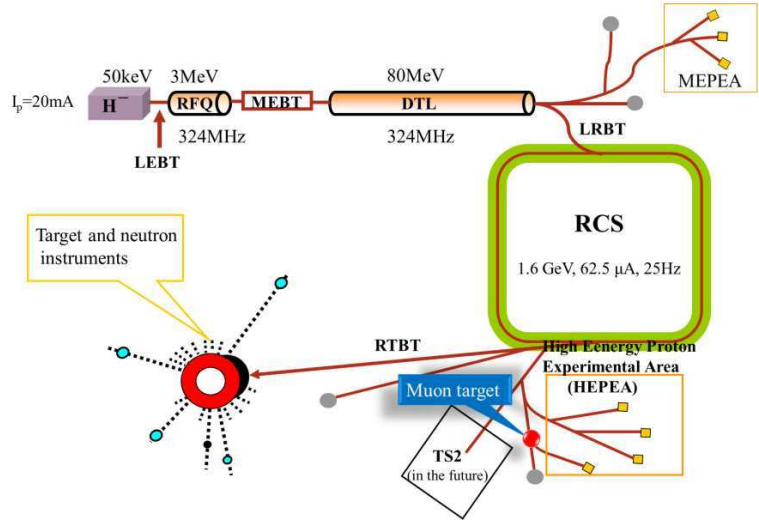}
\caption{(color online) Schematic of China Spallation Neutron Source. Two areas, medium and high energy proton experimental areas (MEPEA and HEPEA), are arranged at CSNS. The preliminary muon beam is considered at HEPEA. TS2 is the second muon target station to be built in the upgrading phase.}
\label{CSNSlayout}
\end{figure*}

The current highest continuous surface muon intensity in the world is 10$^7\sim 5\times10^8$ $\mu^+$/s at PSI \cite{abela1994,prokscha2008} in Switzerland (proton current: 2.2 mA). The highest pulsed surface muon intensity is 10$^7$ $\mu^+$/s  at J-PARC \cite{miyake2006status,miyake2009} in Japan (proton current: 333 $\mu$A). Decay muon rates at PSI and J-PARC are 10$^8$ $\mu^+$/s and 10$^7$ $\mu^+$/s, respectively. High-intensity muon beams can extend muon applications to many subjects such as searching for muon rare decays in particle physics, nuclear muon capture in nuclear physics and ultra slow muon spin rotation in nanomaterials. There is a strong requirement for high intensity muon beams. The normal capture system uses quadrupoles separated outside the target to collection muons and pions at a small solid angle ($<$ 1 sr) acceptance. In the past decade, several designs have been proposed to obtain large acceptance to overcome the limitation caused by the traditional muon capture method. PSI developed a hybird-type acceptance channel \cite{prokscha2008} at the $\mu$E4 beam line using two normal-conducting solenoids with a solid-angle acceptance of $\Delta\Omega\sim$135 msr. KEK \cite{miyadera2006} constructed axial-focusing superconducting coils at Dai Omega delivering up to 2.5 $\times$10$^{5}$ $\mu^{+}$/s at a proton current of 1 $\mu$A with $\Delta\Omega\sim$1 sr. Both of these two beam lines put the target outside the solenoids. RCNP \cite{cook2013first,yoshida2011} proposed a novel surface muon collecting system by placing the target inside a solenoid of 3.5 T. The muon transport system in this institute is upstream of the capture system. In our study, we also consider a superconducting solenoid for the pion-muon capture system with the inner target based on the CSNS proton accelerators \cite{wei2009}. The following solenoid transport system is downstream of capture system. Superconducting solenoids can capture and transport both muons and pions. They are effective not only in realizing a large solid-angle acceptance capture system but also in producing a strong magnetic field of several Teslas with low power consumption \cite{miyadera2006}. 
 
CSNS is a large accelerator-based facility. It produces intense pulsed neutrons by 1.6 GeV protons bombarding a solid tungsten target. The  pulse repetition rate of the protons is 25 Hz. The first-period project (phase I) of CSNS will be finished in 2018. The CSNS complex contains a high beam power proton accelerator, a neutron target station and neutron scattering spectrometers, etc, as shown in Fig. \ref{CSNSlayout}. The proton power of CSNS phase I is 100 kW and can be upgraded to 500 kW in phase II. 
The intensity of each proton pulse is 1.88$\times$10$^{13}$/s, and the effective neutron flux is expected to 2$\times$10$^{16}$ $cm^{-2}s^{-1}$ \cite{wei2009cpc}. CSNS will make a great contribution to the development of many disciplines, such as particle physics, nanoscience, biomedical science, energy and so on. Besides the spallation neutron source, CSNS can provide proton sources for muon beams which will be constructed in the high energy proton experimental area (HEPEA) (Fig. \ref{CSNSlayout}). Research on the muon target and normal beam line at CSNS have been started \cite{wenzhen2012,jing2012,tangjy2010,xu2013}, showing that surface muon intensity of the order of 10$^5$/s at a proton current of 2.5 $\mu$A can be achieved. We present here an advanced large-acceptance channel for pions and muons using superconducting solenoids. The production of muons and pions in the capture system and the beam parameters at the exit of the transport system are analysed by Geant4 \cite{geant1,geant2} (Geant4.9.4.p01) and G4beamline \cite{G4beamline}, respectively.  

\begin{center}
\includegraphics[width=7 cm]{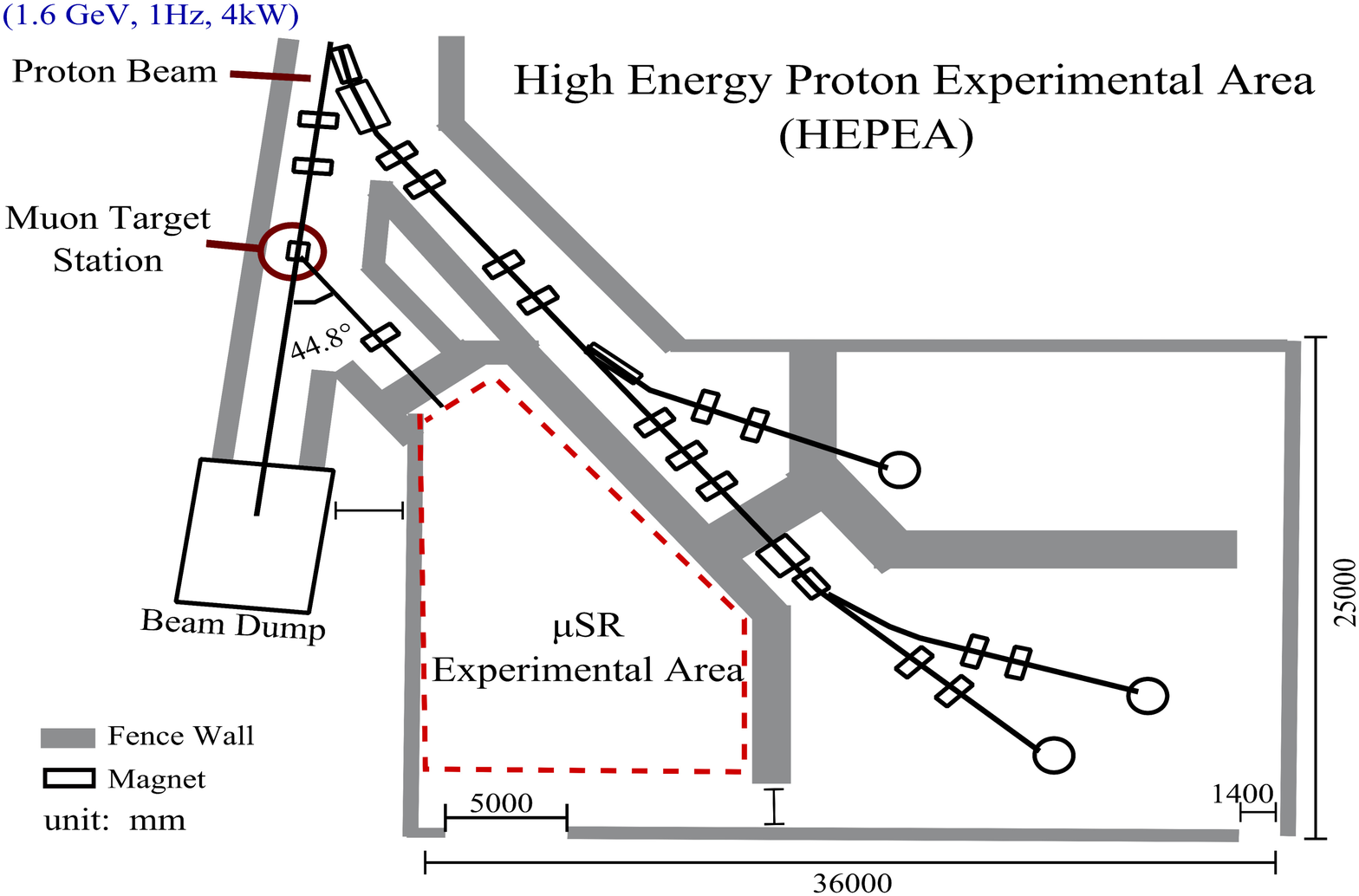}
\figcaption{\label{HEPEA}(color online) Layout of the high intensity proton experimental area at CSNS. The muon experimental area is indicated by the dashed line.}
\end{center}
\section{\label{sec2} A large solid angle capture system for muons and pions using superconducting solenoids}
Figure \ref{HEPEA} shows the layout of the HEPEA area at CSNS, where about 4\% of the proton beam extracted from Rapid Cycling Synchrotron (RCS) is to be injected by halo scraping technique \cite{tang2008beam}. The $\mu$SR instruments will be driven by the proton beam of 4 kW and 2.5 $\mu$A with one repetition rate. Each pulse contains two bunches separated by about 400 ns and the length of each bunch is 70 ns \cite{jing2010}. The beam line of secondary particles produced by protons bombarding the muon target has a 44.8-degree deflection to the proton momentum, and then goes to the $\mu$SR Experiment Area in HEPEA.

\begin{center}
\includegraphics[width=7 cm]{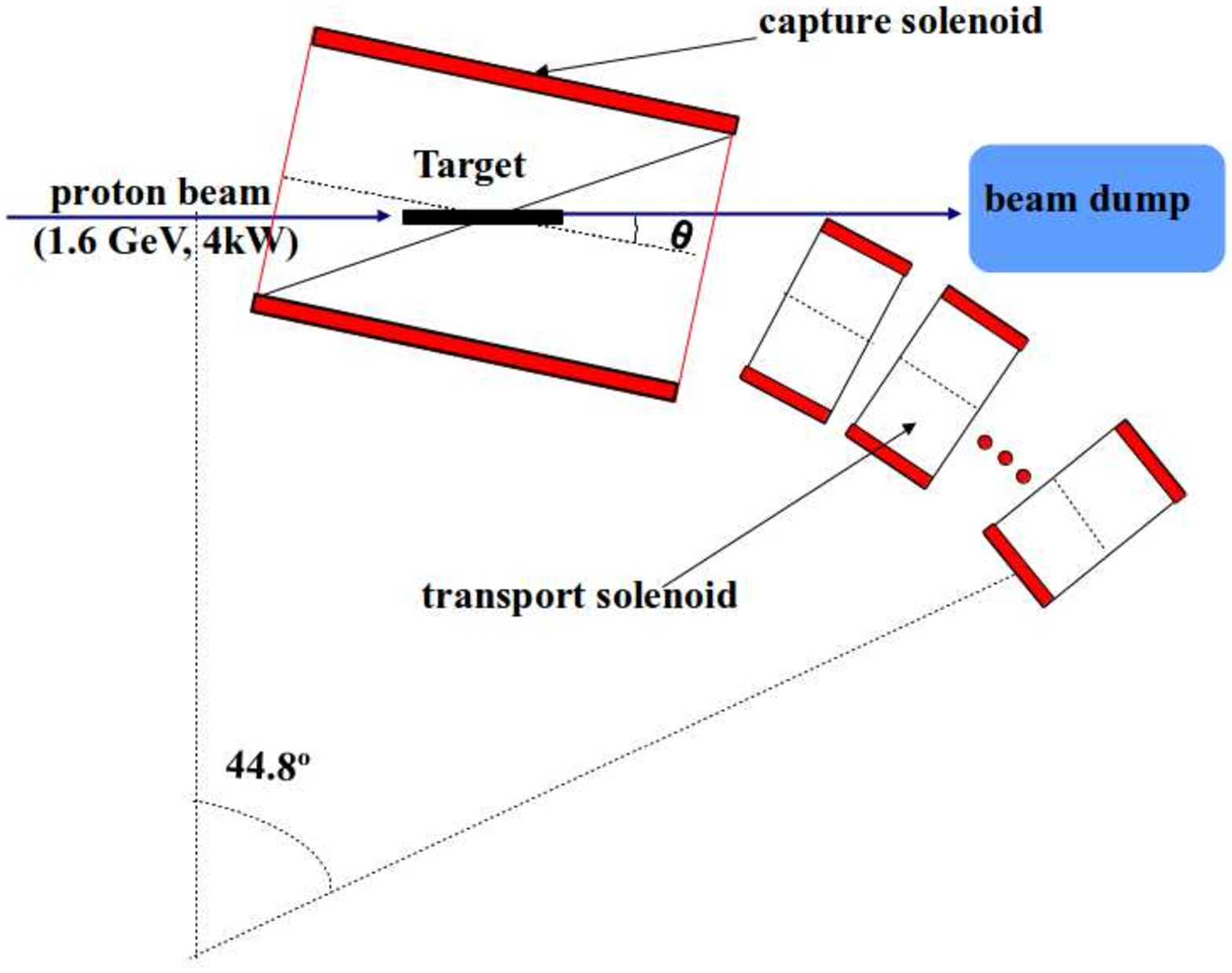}
\figcaption{\label{Target}(color online) Sketch of the inner-target capture system. The target direction is parallel to the proton momentum. $\theta$ is the angle between the solenoid center axis and the proton beam momentum.}
\end{center}
\begin{center}
\includegraphics[width=7 cm]{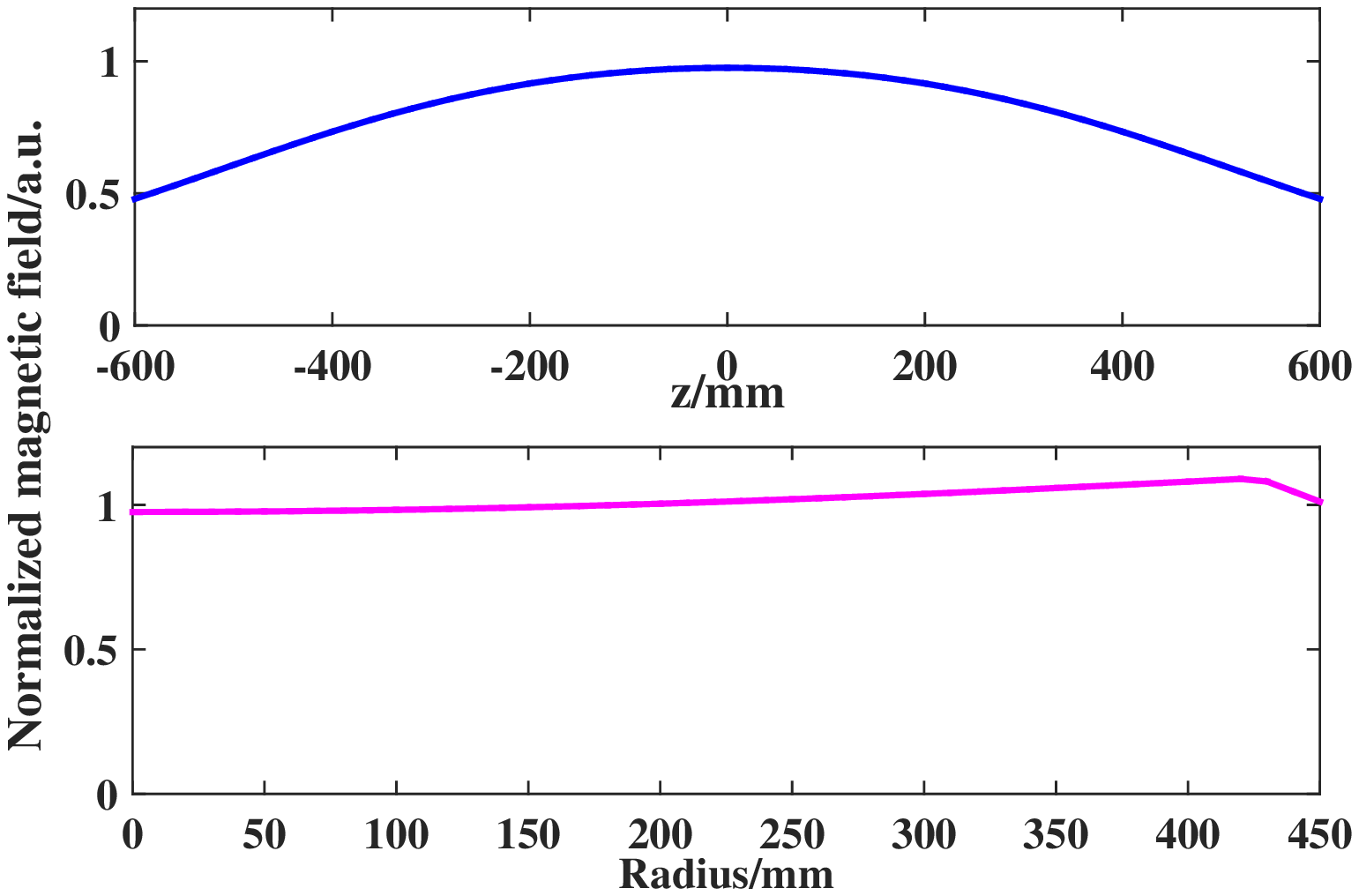}
\figcaption{\label{Magfield}(color online) Normalized magnetic field distribution along the central axis (top) and the radial direction at the median section (bottom) in the capture solenoid in Geant4 setup. The z axis is along the central axis of the solenoid.}
\end{center}

Liu \cite{liu2012muon} has simulated the geometry and material of the target for the normal muon source, and proved by calculation that energy deposition would not cause a great damage to the graphite target. In the current study, a long cylindrical target with dimensions of 40 mm in diameter and 400 mm in length was simulated to produce intense secondary particles. This target is put inside a solenoid with a radius of 450 mm. A solenoid of this radius was chosen to capture more muons and could contain the radiation shields \cite{xiao2014}; the length of the capture solenoid is 1000 mm. Based on the layout of HEPEA, the following muon beam line should be at a 44.8$^\circ$  with respect to the proton beam. The capture and transport system should have the same bending angle to let the beam go smoothly to the muon experimental area. A sketch of the capture system is shown in Fig. \ref{Target}.  Normalized magnetic field distributions along the solenoid axis and radial direction in the median section are shown in Fig. \ref{Magfield}. The parameters of the proton beams used in the simulation are as follows: energy (1.6 GeV), Gaussian function of particles distribution in space ($\sigma_x$=$\sigma_y$=5.732 mm), Gaussian function of angular dispersion distribution ($\sigma_{Xp}$=$\sigma_{Yp}$=14.13 mrad), repetition frequency (1 Hz, 2 bunches/pulse), and beam flux (1.88$\times$10$^{13}$/pps).
\begin{center}
\includegraphics[width=7 cm]{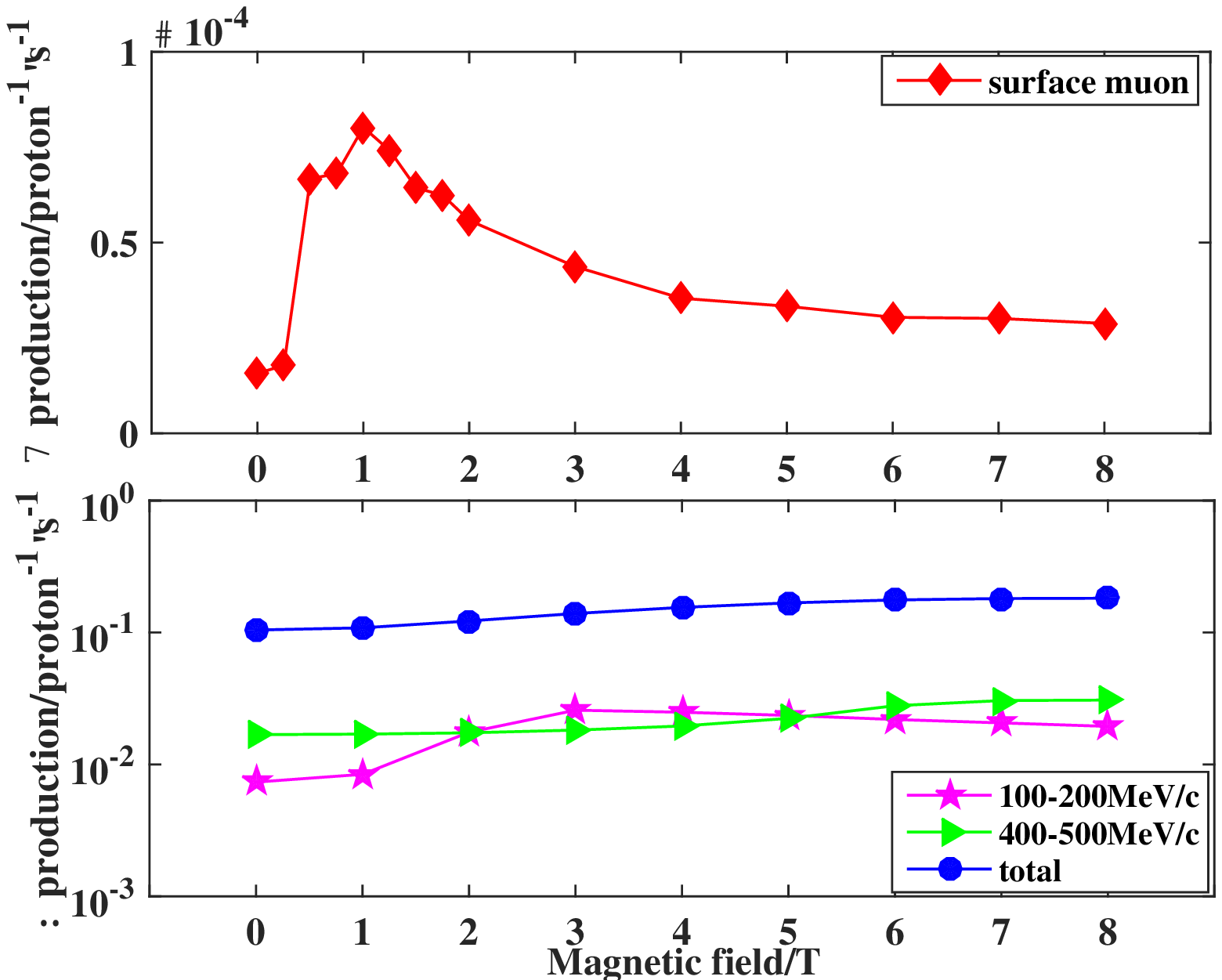}
\figcaption{\label{pmB}(color online) Capture magnetic field dependence of production of surface muons (top) and pions (bottom). The "total" in the bottom figure means all pions with full momenta. The number of primary proton events in the Geant4 simulations is 4$\times$10$^{8}$. }
\end{center}

The magnetic field of the capture solenoid can influence the production of the secondary particles. To find the optimal field to collect more particles, we changed the strength of the magnetic field from 0 to 8 T with a step size of 1 T. Fig. \ref{pmB} shows the production of surface muons (25-30 MeV/c), medium-energy pions (100-200 MeV/c), high-energy pions (400-500 MeV/c) and total pions with full momenta at the downstream exit of the capture solenoid where we put a virtual detector to detect secondary particles. The surface muon production has a significant increase at low magnetic field of the capture solenoid and reaches a peak at about 1 T. It goes down with the increase of the magnetic field and keeps almost the same after the capture magnetic field $>$ 5 T. This production with the capture magnetic field is larger than that in zero field. The capture solenoid with 1 T is the most appropriate collection system for surface muons. For the full-energy and high-energy pions, the production shows a slight increase as the magnetic field goes up. The medium-energy pion production reaches its peak at about 3 T, and then decreases very slowly compared with the peak. The application of the magnetic field at the capture solenoid is an advantage for the collection of secondary particles. The lower the energy of the particles, the smaller the optimal magnetic field is.

\begin{center}
\includegraphics[width=7 cm]{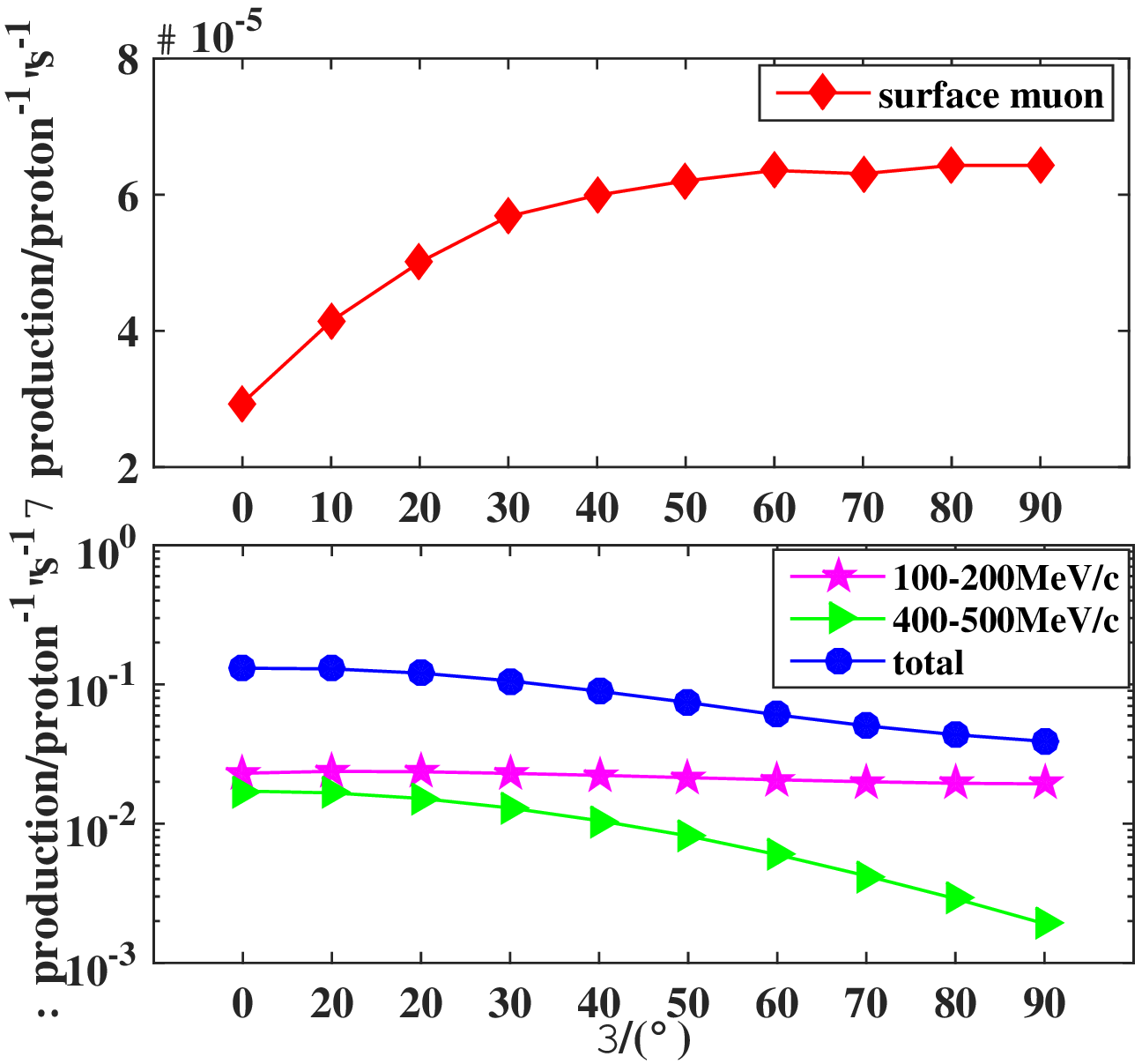}
\figcaption{\label{pmAn}(color online) Production of surface muons (top) and pions (bottom) of polar angle $\theta$ from 0 to 90$^\circ$ with a maximum field magnitude of 3 T at the capture solenoid center. The "total" in the bottom figure means all pions with full momenta. The primary proton event in Geant4 simulations is 4$\times$10$^{8}$. The target is placed along the central axis of the capture solenoid.}
\end{center}

After bombarding the target, most of the incoming protons do not generate secondary particles but go directly through the capture system. The following expression defines the magnetic rigidity of an ideal particle beam with a momentum of $P_{0}$ \cite{liuzp2005}:   
\begin{equation}
(B\rho)_0=P_0/|qe|.
\label{equa1}
\end{equation}    
where $qe$ is the charge of a proton, $\rho$ is the curvature radius and $B$ is the magnetic field. At CSNS the initial proton momentum is 2358 MeV/c and will be $\sim$2200 MeV/c after bombarding the long graphite target. According to equation (\ref{equa1}) when B=3 T, $\rho$ $\sim$2.5 m. The proton beam does not have an obvious deflection in the capture solenoid when the magnetic field is low. Due to the layout of HEPEA (Fig. \ref{HEPEA}), the capture solenoid and the transport solenoids have a 44.8-degree deflection with respect to the proton beam direction. We calculated the production of surface muons and pions at the downstream exit of the capture solenoid of different angles ($\theta$=0$\sim$90$^\circ$) between the solenoid axis and the proton beam momentum (Fig. \ref{pmAn}). The surface muon production ratio has a significant increase when $\theta$ is from 0 to 50$^\circ$ and then goes up slowly. Production of pions with full momenta has a slight decrease after the $\theta$ increase, the medium-energy pion production stays almost the same, but for high-energy pions their production has an obvious decrease when $\theta$ is larger than 20$^\circ$. This may be because most high-energy pions have a large curvature radius and are killed by the inner side of the capture solenoid in the simulation when the superconducting solenoid is not parallel to the proton momentum. Finally, $\theta$ is chosen as 22.8$^\circ$ which will be better to forward the rest protons to the beam dump \cite{xiao2014} and to reduce the proton radiation damage.  
\begin{figure*}[ht]
\centering
\includegraphics[width=12 cm]{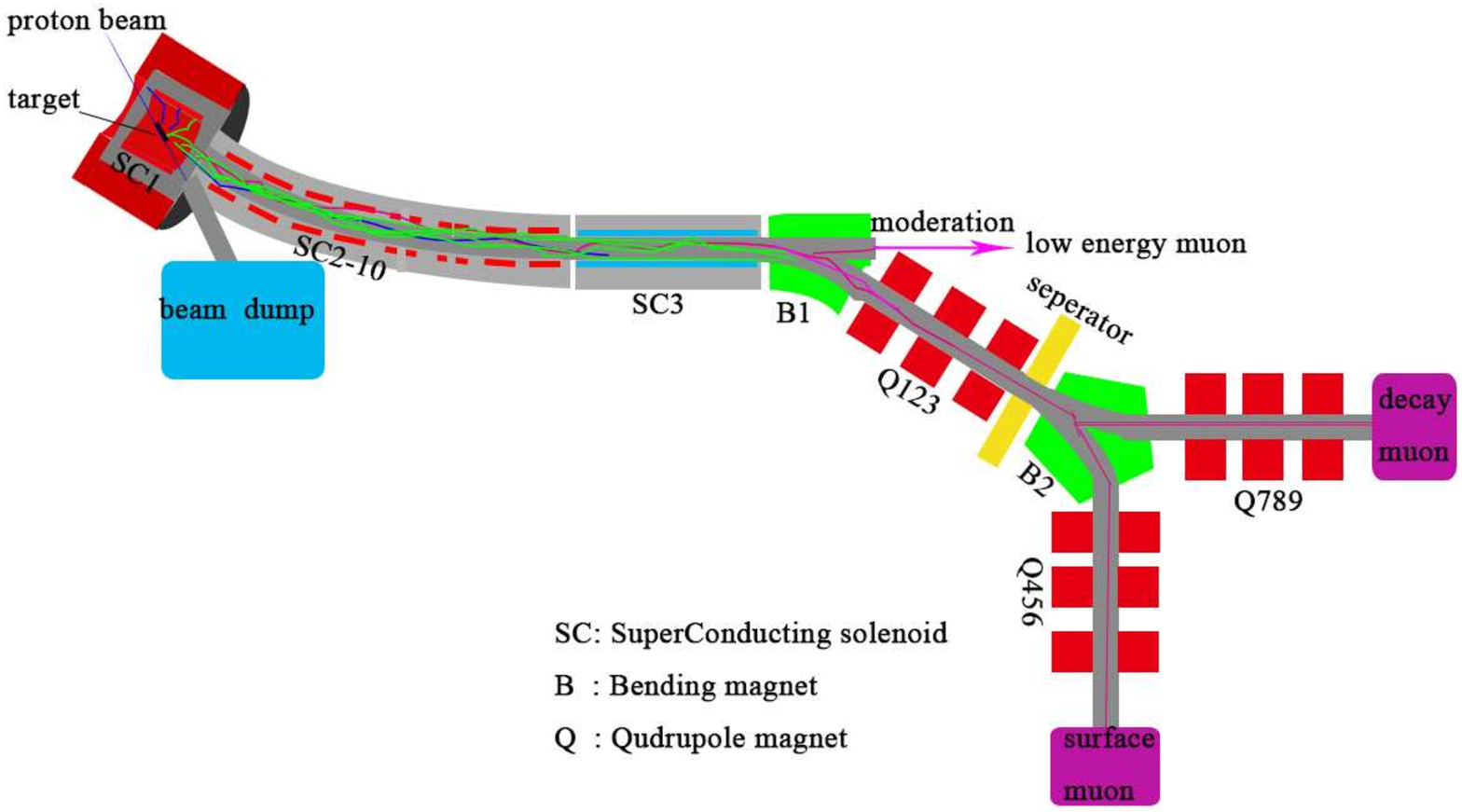}\\
\figcaption{(color online) Preliminary schematic of pion and muon beam lines ($\pi-\mu$ line). SC1 is the capture solenoid described in section 2. SC2-10 contains ten short superconducting solenoids with lengths of 200 mm. SC3 is the decay solenoid.}
\label{Solelayout}
\end{figure*} 
\section{\label{sec3} Design of the surface and decay muon transport system}
\begin{figure*}[ht]
\centering
\includegraphics[width=13 cm]{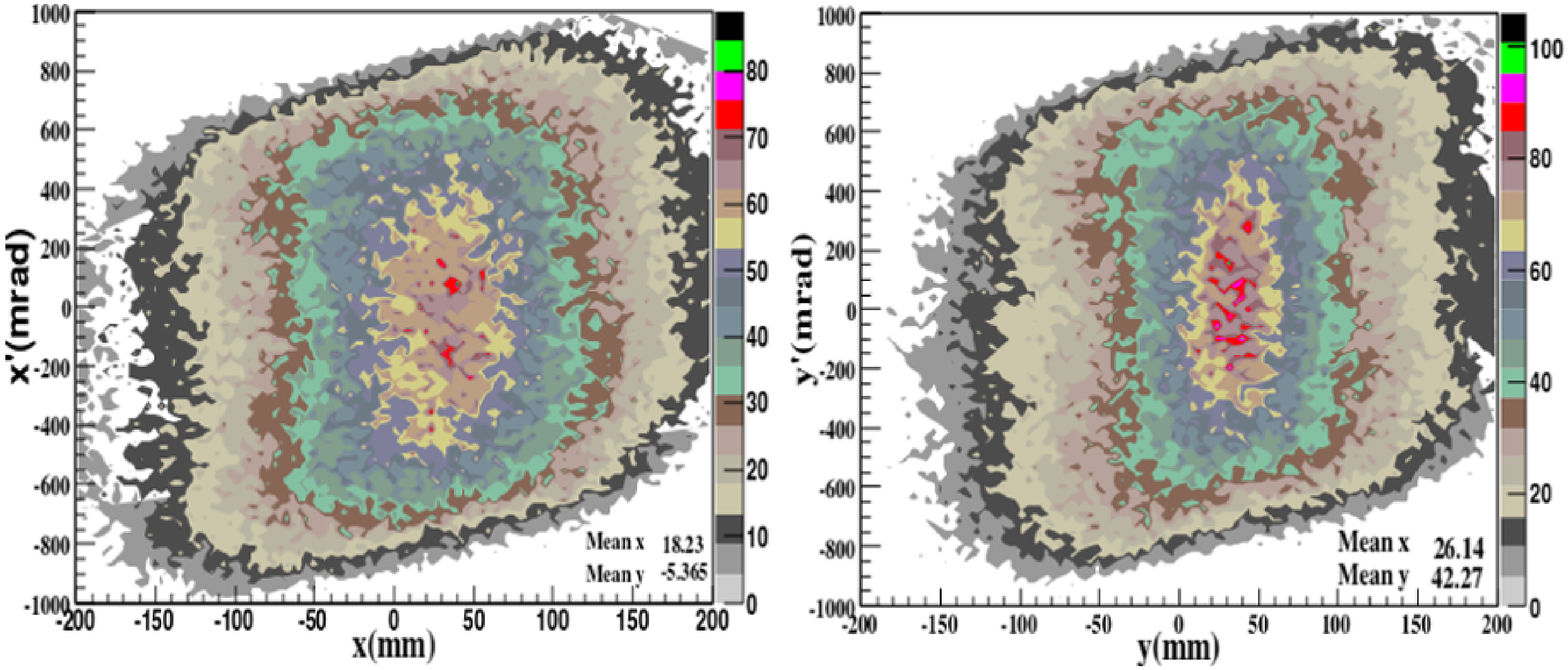}
\figcaption{\label{pionPhaseSpace}(color online) Phase space distribution of decay muons on the horizontal (left) and vertical (right) axis at the exit of the decay solenoids. The labels 'x' and 'x$^\prime$' in the left figure mean the horizontal position and angular divergence of decay muons, 'y' and 'y$^\prime$' in the right figure represent the vertical position and angular divergence of the decay muons. The number of the primary proton events in Geant4 simulations is 4$\times$10$^{8}$. }
\label{pionPhase}
\end{figure*}

Secondary particles captured in the superconducting solenoid will be transported and focused to the following beam elements. Ten short superconducting solenoids with maximum field magnitude of 2 T will be used to focus and bend muon and pion beams to the following beam elements.  Each axis of the ten solenoids has a 2-degree deflection  with a curved radius of 3 m with respect to the corresponding direction of the beam line. The mean decay length of pions \cite{nagamine2003} whose lifetime is 26 ns is as follows:
 \begin{equation}
 L_{\pi}(cm)=c\beta\gamma_{\pi}\tau_{\pi}=5.593\times p_{\pi}(MeV/c). 
 \end{equation}
where $\tau_{\pi}$ is the mean lifetime of pions at rest, and $p_{\pi}$ is the momentum of pions. For the medium-energy pions (100$\sim$200 MeV), $L_{\pi}$ is about 5$\sim$11 m. A decay channel is necessary in a beam line to achieve more decay muons at the experiment areas. Fig. \ref{Solelayout} gives the preliminary schematic of the pion-muon capture and transport system. The solenoid S3 with length of 2 m is used as the decay channel with maximum field magnitude of 2 T. The decay channel is followed by a bending magnet (B1) which can bend the beam as well as select particles with the momentum of interest (the principle is as equation (\ref{equa1})). After the bending magnet, surface and decay muons will be transported to the surface and decay muon terminal experiment areas, respectively. These muons can also go straight to a moderation area after B1. They can be moderated to low energy muons (eV$\sim$keV) using the cryogenic moderator method \cite{bakule2004} or laser ionization method \cite{miyake2013}.
These muon moderation methods have broadened the $\mu$SR technique to the areas of nano-materials and multi-layer compounds \cite{prokscha2004,morenzoni2004nano,eshchenko2009low}. 

\begin{table*}[t]
\centering
\caption{Emittance and intensity of surface and decay muon beams at the exit of the decay solenoids. The number of primary proton events  in Geant4 simulations is 4$\times$10$^{8}$. The production in the table is calculated according to the proton flux of the CSNS (1.88$\times$10$^{13}$/s).}
\begin{tabular}{ccrrccc}
\hline 
Phase plane & Type & \multicolumn{2}{c}{$\epsilon_{RMS}$($\pi$ mm $\cdot$ mrad)} & & \multicolumn{2}{c}{Intensity ($\mu^+$/s)} \\ 
\cline{3-4}\cline{6-7}
 &  & Surface $\mu^+$ & Decay $\mu^+$ &  & Surface $\mu^+$ & Decay $\mu^+$ \\ 
\hline
x-x' & Beam core & 12358 & 10000  &  & 1.67$\times$10$^8$ & 1.10$\times$10$^9$ \\ 
     & 80\%      & 79701 & 120000 &  & 7.93$\times$10$^8$ & 8.70$\times$10$^9$ \\ 
y-y' & Beam core & 7500  & 70000  &  & 1.95$\times$10$^8$ & 1.06$\times$10$^9$ \\ 
     & 80\%      & 51000 & 114540 &  & 7.06$\times$10$^8$ & 8.95$\times$10$^9$ \\ 
\hline 
\end{tabular} 
\label{table1}
\end{table*}  
  
The decay muon intensities and phase space distributions at the exit of SC3 were calculated by G4beamline as a first step with the following magnetic field parameters: B$_{SC1}$=3 T, B$_{SC2}$=B$_{SC3}$=2 T. Fig. \ref{pionPhase} shows the phase space of decay muons with momentum of 100 MeV/c and momentum bite ($\pm\delta$p/p) of 5\%. The surface muon phase space can be seen in ref \cite{xiao2014} for details. From Fig. \ref{pionPhaseSpace}, the divergence of the decay muons is very large, about 5$\sim$6 times larger than that designed by the normal quadrupole transport system \cite{xu2013}. Using the phase ellipse fitting method \cite{liuzp2005}, we calculated emittances ($\epsilon_{RMS}$) and intensities of beam core and 80\% beam ratios of surface and decay muons, as shown in Table \ref{table1}. The intensities of the beam core of surface muons and decay muons at the exit of the decay solenoids reached 10$^8$/s and 10$^9$/s with a proton current of 2.5 $\mu$A, respectively. The intensity is about two or three orders larger than that in normal beam line (10$^5\sim$10$^6$ $\mu^+$/s). The cryogenic moderator method efficiency is about 10$^{-4}$ or 10$^{-5}$ \cite{morenzoni1998}, so the high intensity muon beam is necessary for the low energy muon beam; in this $\pi-\mu$ line after moderation the low energy muon intensity can still be 10$^3$ or 10$^4$ $\mu^+$/s. The ellipse Twiss (CS) parameters of the muon beam obtained at the exit of SC3 will be used as original beam parameters in the future to fit to optimal small spot beams by matrix multiplication programs for beam optic designs.

\section{\label{sec4}Summary and prospects}
Using 4\% of the proton beam extracted from RCS at CSNS as an initial driven beam source, we have designed the muon and pion channel in the HEPEA area. Firstly, a superconducting solenoid with an inner graphite target was proposed as the muon collection system. The muon and pion intensities of different magnetic fields and angles of the capture solenoid with respect to the proton beam were calculated by Geant4. The angle between the solenoid axis and proton beam direction was designated as 22.8$^\circ$. The superconducting solenoid system was designed by G4beamline based on the project of HEPEA plans, and phase space distributions and intensities of muons at the exit of decay solenoids were investigated. Finally, it was found that the surface muon rate could reach 10$^8$ $\mu^+$/s and the decay muon rate achieve 10$^9$ $\mu^+$/s. These results suggest that the superconducting collection and capture system can improve the muon intensity. 

These high-flux muon beams will be optimized to achieve the required beams with a small beam spot in the terminal experiment area. More detailed characteristics of the muon beams will discussed in the future, including the polarization, the beam envelopes of the surface and decay muons and reduction of the contamination from low energy protons and positrons by a separator. A cryogenic moderation method using a wide-band-gap van der Waals solid gas, similar to that used by PSI, will be adopted to producing the low energy muons. The $\pi-\mu$ line at CSNS will provide three kinds of $\mu$SR experiment areas and will be an innovative muon beam line. 

\section{\label{sec5}Acknowledgements}
The authors are grateful to Elvezio Morenzoni and Thomas Prokscha at PSI for their valuable suggestions in the CSNS muon beam line design, and also thank Wei Kong for useful discussions.
\end{multicols}
\end{CJK*}

\clearpage

\begin{thebibliography}{10}

\begin{multicols}{2}
\bibitem{Yaouanc2011}
A~Yaouanc and P~D R{\'e}otier.
\newblock {\em Muon spin rotation, relaxation, and resonance: applications to
  condensed matter}.
\newblock Number 147. Oxford University Press, 2011.

\bibitem{nagamine2003}
K~Nagamine.
\newblock {\em Introductory muon science}.
\newblock Cambridge University Press, 2003.

\bibitem{brewer1981}
Jess~H Brewer.
\newblock $\mu$+ sr with surface muon beams.
\newblock {\em Hyperfine Interactions}, 8(4):831--834, 1981.

\bibitem{abela1994}
R~Abela, C~Baines, X~Donath, D~Herlach, D~Maden, I~D Reid, D~Renker, G~Solt,
  and U~Zimmermann.
\newblock The $\mu$sr facilities at psi.
\newblock {\em Hyperfine Interactions}, 87(1):1105--1110, 1994.

\bibitem{prokscha2008}
T~Prokscha, E~Morenzoni, K~Deiters, F~Foroughi, D~George, R~Kobler, A~Suter,
  and V~Vrankovic.
\newblock The new $\mu$e4 beam at psi: A hybrid-type large acceptance channel
  for the generation of a high intensity surface-muon beam.
\newblock {\em Nuclear Instruments and Methods in Physics Research Section A:
  Accelerators, Spectrometers, Detectors and Associated Equipment},
  595(2):317--331, 2008.

\bibitem{miyake2006status}
Y~Miyake, K~Nishiyama, N~Kawamura, S~Makimura, P~Strasser, K~Shimomura,
  JL~Beveridge, R~Kadono, K~Fukuchi, N~Sato, et~al.
\newblock Status of j-parc muon science facility at the year of 2005.
\newblock {\em Physica B: Condensed Matter}, 374:484--487, 2006.

\bibitem{miyake2009}
Y~Miyake, K~Nishiyama, N~Kawamura, P~Strasser, S~Makimura, A~Koda, K~Shimomura,
  H~Fujimori, K~Nakahara, R~Kadono, et~al.
\newblock J-parc muon source, muse.
\newblock {\em Nuclear Instruments and Methods in Physics Research Section A:
  Accelerators, Spectrometers, Detectors and Associated Equipment},
  600(1):22--24, 2009.

\bibitem{miyadera2006}
H~Miyadera, K~Nagamine, K~Shimomura, K~Nishiyama, K~Fukuchi, and K~Ishida.
\newblock Design, construction and performance of dai omega, a large
  solid-angle axial-focusing superconducting surface-muon channel.
\newblock {\em Nuclear Instruments and Methods in Physics Research Section A:
  Accelerators, Spectrometers, Detectors and Associated Equipment},
  569(3):713--726, 2006.

\bibitem{cook2013first}
S~Cook, R~D'Arcy, M~Fukuda, K~Hatanaka, Y~Hino, Y~Kuno, M~Lancaster, Y~Mori,
  TH~Nam, T~Ogitsu, et~al.
\newblock First measurements of muon production rate using a novel pion capture
  system at music.
\newblock In {\em Journal of Physics: Conference Series}, volume 408, page
  012079. IOP Publishing, 2013.

\bibitem{yoshida2011}
M~Yoshida, M~Fukuda, K~Hatanaka, Y~Kuno, T~Ogitsu, A~Sato, and A~Yamamoto.
\newblock Superconducting solenoid magnets for the music project.
\newblock {\em Applied Superconductivity, IEEE Transactions on},
  21(3):1752--1755, 2011.

\bibitem{wei2009}
J~Wei, H~S Chen, Y~W Chen, Y~B Chen, Y~L Chi, C~D Deng, H~y Dong, L~Dong, S~X
  Fang, J~Feng, et~al.
\newblock China spallation neutron source: design, r\&d, and outlook.
\newblock {\em Nuclear Instruments and Methods in Physics Research Section A:
  Accelerators, Spectrometers, Detectors and Associated Equipment},
  600(1):10--13, 2009.

\bibitem{wei2009cpc}
J~Wei, S~N Fu, J~Y Tang, J~Z Tao, D~S Wang, F~W Wang, and S~Wang.
\newblock China spallation neutron source-an overview of application prospects.
\newblock {\em Chinese Physics C}, 33(11):1033, 2009.

\bibitem{wenzhen2012}
W~Z Xu, Y~F Liu, and B~J Ye.
\newblock Simulation and design of tentative muon source based on csns.
\newblock {\em Plasma Science and Technology}, 14(6):469, 2012.

\bibitem{jing2012}
H~T Jing, C~Meng, J~Y Tang, B~J Ye, and J~L Sun.
\newblock Production target and muon collection studies for an experimental
  muon source at csns.
\newblock {\em Nuclear Instruments and Methods in Physics Research Section A:
  Accelerators, Spectrometers, Detectors and Associated Equipment},
  684:109--116, 2012.

\bibitem{tangjy2010}
J~Y Tang, S~N Fu, H~T Jing, H~Q Tang, J~Wei, and H~H Xia.
\newblock Proposal for muon and white neutron sources at csns.
\newblock {\em Chinese Physics C}, 34(1):121, 2010.

\bibitem{xu2013}
W~Z Xu.
\newblock {\em Design of Surface Muon Source Based on Spallation Neutron Source
  and Study of the Related Simulated Techniques}.
\newblock PhD thesis.

\bibitem{geant1}
S~Agostinelli, J~Allison, K~al Amako, J~Apostolakis, H~Araujo, P~Arce, M~Asai,
  D~Axen, S~Banerjee, G~Barrand, et~al.
\newblock Geant4—a simulation toolkit.
\newblock {\em Nuclear instruments and methods in physics research section A:
  Accelerators, Spectrometers, Detectors and Associated Equipment},
  506(3):250--303, 2003.

\bibitem{geant2}
J~Allison, K~Amako, J~Apostolakis, HAAH Araujo, P~Arce Dubois, MAAM Asai, GABG
  Barrand, RACR Capra, SACS Chauvie, RACR Chytracek, et~al.
\newblock Geant4 developments and applications.
\newblock {\em Nuclear Science, IEEE Transactions on}, 53(1):270--278, 2006.

\bibitem{G4beamline}
\url{http://www.muonsinternal.com/muons3/G4beamline}.

\bibitem{tang2008beam}
J~Y Tang, G~H Wei, C~Zhang, J~Qiu, L~Lin, and J~Wei.
\newblock Beam preparation for the injection into csns rcs.
\newblock {\em Proc. of HB2008}, 2008.

\bibitem{jing2010}
HT~Jing, JY~Tang, HQ~Tang, HH~Xia, TJ~Liang, ZY~Zhou, QP~Zhong, and XC~Ruan.
\newblock Studies of back-streaming white neutrons at csns.
\newblock {\em Nuclear Instruments and Methods in Physics Research Section A:
  Accelerators, Spectrometers, Detectors and Associated Equipment},
  621(1):91--96, 2010.

\bibitem{liu2012muon}
Y~F Liu, W~Z Xu, Z~Q Tan, Y~Liang, W~Kong, and B~J Ye.
\newblock Design and optimization for experimental muon source at csns.
\newblock {\em SCIENTIA SINICA Physica, Mechanica \& Astronomica},
  42(11):1204--1211, 2012.

\bibitem{xiao2014}
R~Xiao, Y~F Liu, W~Z Xu, Z~Q Tan, B~Cheng, W~Kong, and B~J Ye.
\newblock Study on a new large solid angle capture system for surface muon
  using superconducting solenoids.
\newblock {\em Nuclear Physics Review}, 31(4):468--474, 2014.

\bibitem{liuzp2005}
Z~P Liu.
\newblock {\em Beam Optics}.
\newblock Press of University of Science and Technology of China, 2005.

\bibitem{bakule2004}
P~Bakule and E~Morenzoni.
\newblock Generation and applications of slow polarized muons.
\newblock {\em Contemporary Physics}, 45(3):203--225, 2004.

\bibitem{miyake2013}
Y~Miyake, K~Ikedo, R~Kadono, E~Torikai, et~al.
\newblock {\em Hyperfine interactions}, 216, 2013.

\bibitem{prokscha2004}
T~Prokscha, E~Morenzoni, A~Suter, R~Khasanov, H~Luetkens, D~Eshchenko,
  N~Garifianov, EM~Forgan, H~Keller, J~Litterst, et~al.
\newblock Thin film, near-surface and multi-layer investigations by low-energy
  $\mu$+ sr.
\newblock {\em Hyperfine interactions}, 159(1-4):227--234, 2004.

\bibitem{morenzoni2004nano}
E~Morenzoni, T~Prokscha, A~Suter, H~Luetkens, and R~Khasanov.
\newblock Nano-scale thin film investigations with slow polarized muons.
\newblock {\em Journal of Physics: Condensed Matter}, 16(40):S4583, 2004.

\bibitem{eshchenko2009low}
DG~Eshchenko, VG~Storchak, E~Morenzoni, T~Prokscha, A~Suter, X~Liu, and
  JK~Furdyna.
\newblock Low energy $\mu$sr studies of semiconductor interfaces.
\newblock {\em Physica B: Condensed Matter}, 404(5):873--875, 2009.

\bibitem{morenzoni1998}
E~Morenzoni.
\newblock Physics and applications of low energy muons.
\newblock {\em Muon Science: Muons in Physics, Chemistry and Materials (Bristol
  and Philadelphia, 1999), SL Lee, SH Kilcoyne, and R. Cywinski, Eds},
  51:343--404, 1998.

\end{multicols}
\end{thebibliography}
\end{document}